# INTEGRAL and Multiwavelength Observations of the Blazar Mrk 421 during an Active Phase

G. G. Lichti[1], E. Bottacini[1], P. Charlot[2], W. Collmar[1], D. Horan[3], A. von Kienlin[1], A. Lähteenmäki[4], K. Nilsson[5], D. Petry[1], A. Sillanpää[5], M. Tornikoski[4], T. Weekes[3]

[1]Max-Planck-Institut für extraterrestrische Physik, Giessenbachstrasse, 85748 Garching, Germany;
[2]Observatoire de Bordeaux, Floirac, France;
[3]Fred Lawrence Whipple Observatory, Amado, USA;
[4]TKK / Metsähovi Radio Observatory, Kylmälä, Finland;
[5]Tuorla Observatory, Piikkiö, Finland

**Abstract.** A ToO observation of the TeV-emitting blazar Mrk 421 with INTEGRAL was triggered in June 2006 by an increase of the RXTE count rate to more than 30 mCrab. The source was then observed with all INTEGRAL instruments with the exception of the spectrometer SPI for a total exposure of 829 ks. During this time several outbursts were observed by IBIS and JEM-X. Multiwavelength observations were immediately triggered and the source was observed at radio, optical and X-ray wavelengths up to TeV energies. The data obtained during these observations are analysed with respect to spectral evolution and correlated variability. Preliminary results of the analysis are presented in this poster.

## OBSERVATIONS WITH INTEGRAL

Mrk 421, the 1st AGN detected at TeV energies, became active (>30 mCrab measured by RXTE) in April 2006. This triggered an INTEGRAL ToO observation with the INTEGRAL instruments OMC, JEM-X and IBIS between June 14 – 25 (SPI unfortunately was not operating because of annealing). Mrk 421 was detected by all 3 instruments with a high significance (up to 160σ between 20 – 50 keV). Multiwavelength observations at all wavelengths were initiated and preliminary results are presented here.

In Figure 1 the ISGRI and the OMC lightcurves are shown. Whereas the optical lightcurve measured by OMC scatters around a mean value of ~2.6x10$^{-14}$ erg/(cm² s Å), four strong flares are observed at hard X-rays by ISGRI. The strongest flare is a factor of 4 more intense than the quiescent level of ~4 cts/s. The time interval was split into different phases with either quiescent or active emission. The data within each state were then spectrally investigated.

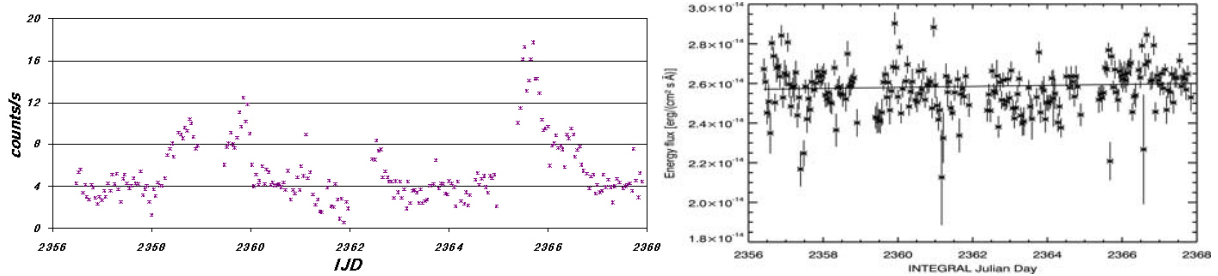

**FIGURE 1.** The ISGRI lightcurve (left; 20-900 keV) and the OMC lightcurve (right) of Mrk 421 in June 2006.

## The Spectral Analysis

Different spectral models (simple power law (PL), broken PL, PL with smooth transition, Band function, PL with exp. cutoff, log-parabolic law) were fitted (XSPEC) to JEM-X and ISGRI data, covering quiescent and flaring states, to find the function which fits the data best. A broken PL could be best fitted in both cases ($\chi^2_{red}$ = ~1.9). In Table 1 the best-fit parameters are given. Apart from A and $\alpha$, the parameters of the 2 states are consistent within errors. The decrease of $\alpha$ in the active state, however, indicates a slight spectral hardening at 3.6$\sigma$. The best-fitted spectra are shown in Figure 2.

**TABLE 1.** The best-fit parameters of the broken power-law model.

| Parameter Type | Quiescent State | Active State |
|---|---|---|
| Normalization Constant A [cts/(s keV)] | 0.365 ± 0.005 | 0.49 ± 0.005 |
| Low-Energy PL Index $\alpha$ | 2.300 ± 0.006 | 2.12 ± 0.05 |
| High-Energy PL Index $\beta$ | 3.0 ± 0.2 | 2.90 ± 0.08 |
| Break Energy [keV] | 45 ± 4 | 41 ± 2 |

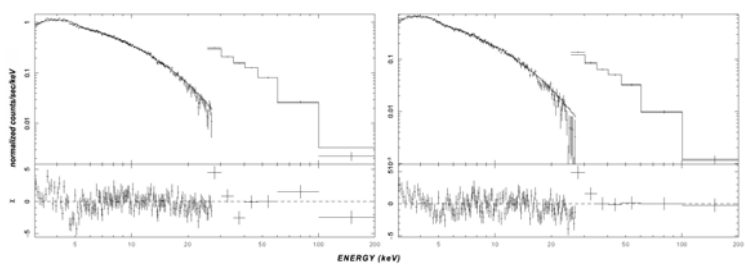

**FIGURE 2.** The normalized counts/(s keV) of the XSPEC fits are shown as a function of energy (left: active state; right: quiescent state). In the bottom panels the residuals of the fits are plotted.

## Multiwavelength Observations and Quasi-Simultaneous $\nu F_\nu$ Spectrum

In parallel to the INTEGRAL observations multiwavelength observations were initiated at radio (Metsähovi and VLBA radiotelescopes), optical (KVA telescope) and TeV energies (Whipple). The data from these observations were averaged over the observation time and then compiled in an energy-density spectrum (Figure 3). They are compared with the theoretical models of Maraschi et al. (1999)[1] and Blazejowski et al. (2005)[2] which were adapted in intensity to the quiet IBIS data and serve only the purpose to guide the eyes.

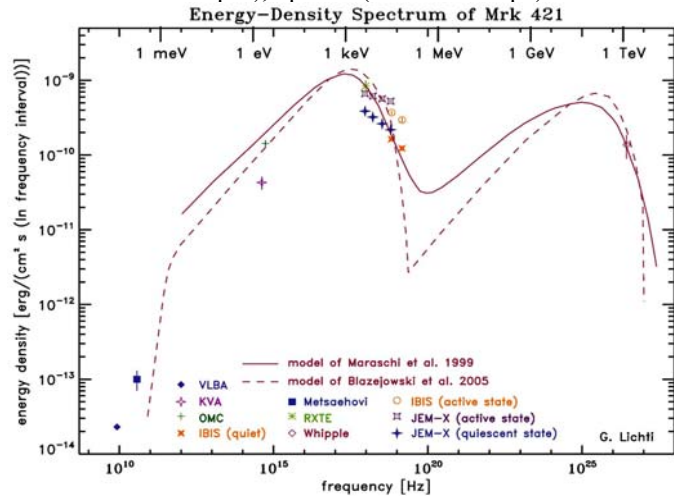

**FIGURE 3.** The $\nu F_\nu$ spectrum of Mrk 421 for quasisimultaneous data. In addition the data for the two states are shown at X-rays and compared with two theoretical models.

## CONCLUSIONS

Both models fit the RXTE, IBIS and Whipple data reasonably well, but the model of Maraschi et al. (1999)[1] predicts higher energy densities at optical wavelengths than observed with OMC. However, the spectra measured by JEM-X and IBIS in both the active and quiescent states are flatter than predicted by the models. Also a slight inconsistency of the inter-instrument calibration between JEM-X and IBIS is obvious.

## ACKNOWLEDGMENTS

The INTEGRAL project was supported by the BMBF via the DLR under the contract number 50.OG.9503.0.